# Supply Chain Network Security Investment Strategies Based on Nonlinear Budget Constraints: The Moderating Roles of Market Share and Attack Risk

Jiajie Cheng[1], Jiaxin Wang[2], Caijiao Li[3], Luxiang Zhang[4], Yusheng Fan[3], Yujie Bao[1] and Wen Zhou[1*]

[1] Nanjing University of Finance & Economics, Nanjing Jiangsu 210023, China
[2] Tianjin University of Science & Technology, Tianjin 300222, China
[3] Fudan University, Shanghai 200433, China
[4] University of California Santa Barbara, Santa Barbara, CA 93106, USA
[5] Tianjin University, Tianjin 300072, China
Corresponding email: Zw_academic08@163.com

**Abstract.** In the context of the rapid development of digital supply chain networks, dealing with the increasing cybersecurity threats and formulating effective security investment strategies to defend against cyberattack risks are the core issues in supply chain management. Cybersecurity investment decision-making is a key strategic task in enterprise supply chain management. Traditional game theory models and linear programming methods make it challenging to deal with complex problems such as multi-party participation in the supply chain, resource constraints, and risk uncertainty, resulting in enterprises facing high risks and uncertainties in the field of cybersecurity. To effectively meet this challenge, this study proposes a nonlinear budget-constrained cybersecurity investment optimization model based on variational inequality and projection shrinkage algorithm. This method simulates the impact of market competition on security investment by introducing market share variables, combining variational inequality and projection shrinkage algorithm to solve the model, and analyzing the effect of different variables such as budget constraints, cyberattack losses, and market share on supply chain network security. In numerical analysis, the model achieved high cybersecurity levels of 0.96 and 0.95 in the experimental scenarios of two retailers and two demand markets, respectively, and the budget constraint analysis revealed the profound impact of budget constraints on cybersecurity investment. Through numerical experiments and comparative analysis, the effectiveness and operability of this method in improving supply chain network security are verified.

# 1 Introduction

With the rapid development of Internet technology and information technology, cybercrime incidents occur frequently, which poses a serious threat to the operation of supply chain networks. Cybercriminals usually use various means to attack corporate supply chain systems, which may lead to key data leakage, production interruptions, and financial losses. At the same time, there is a risk propagation effect due to cross-correlations in various supply chain links. This leads to the fact that any attack on a single node will trigger a chain reaction in the entire supply chain, affecting many areas. For example, in 2023, the large-scale leakage of PayPal users' personal information further highlighted the severity of cybersecurity threats. At the same time, with the rise of new technologies such as generative artificial intelligence, the threshold for cyberattacks has been significantly lowered, and the means of attack have become more sophisticated. The prevalence of zero-day vulnerabilities and destructive hacking has further exacerbated this trend, and global cybercrime losses are expected to hit a record high in 2026. Therefore, formulating a scientific and effective cybersecurity defense strategy is crucial to ensuring data security, production continuity, and economic benefits for supply chain companies.

Affected by budget constraints, retailers must reasonably allocate cybersecurity investments under limited resources to maximize investment benefits. Retailers with large market shares are prone to become the primary targets of cyber-attacks due to their extensive business scale and large customer data. They must conduct effective supply chain cybersecurity research and formulate high-level cybersecurity investment plans that match their market position to ensure stable supply chain operation, prevent production interruptions and data leakage risks, and maintain corporate reputation and customer trust.

For supply chain cybersecurity issues, many studies currently focus on using network equilibrium theory and information-sharing strategies for risk management. For example, Nagurney et al. [2–4] and Qian et al. [5] use Nash equilibrium and partially centralized decision-making methods to solve the problem of information sharing and security investment problem between enterprises. At the same time, Liu et al. [6], Ezhei et al. [7], Solak et al. [8], etc., based on the above work, further expand the scope of research on the impact of security interdependence and investment externalities. These works explain to a certain extent that the methods based on multi-agent games and network equilibrium have become a critical path to solving the current supply chain cybersecurity issues.

Methods based on differential games and stochastic programming can help capture the security interdependence caused by collaborative decision-making and changes in the external environment in the supply chain network and effectively solve the problem of insufficient investment under budget constraints and demand fluctuations. However, existing studies still cannot eliminate the problem of unbalanced security investment allocation from the perspective that companies with larger market shares are more likely to become the primary targets of attack. Li and Xu [9] reduced expected losses through joint decision-making and risk compensation strategies. Cheung et al. [10] and Aldrighetti et al. [11] analyzed the impact of cyber-attack threats and disruption on the IoT supply chain. Sawik [12] and Colajanni et al. [13] improved the investment strategy using recursive linearization and nonlinear budget constraint models. However, in the context of multi-retailer competition, studies by domestic scholars such as Wang Mao [18], Zhang Zijian et al. [21], and Yao Hong [23] have shown that if market share differences and security externalities are ignored, the supply chain network often faces insufficient security protection or imbalanced investment allocation.

Based on the above supply chain network security background, this paper constructs a nonlinear budget-constrained network security investment optimization model based on variational inequality and projection shrinkage algorithm. It introduces key variables, such as market share, to analyze the impact of these variables on supply chain network security strategy. This model provides a scientific analysis tool for retailers to optimize network security investment under different market conditions and theoretical support and decision-making assistance for supply chain network security management in the digital age. It is essential to build a safe and reliable supply chain ecosystem and promote the healthy development of the supply chain. Specifically, this method has the following three main innovations:

- The variational inequality and the projection shrinkage algorithm are organically combined to achieve a fast iterative solution for supply chain network security investment under high-dimensional nonlinear budget constraints, overcoming the applicability limitations of traditional linear programming models in dealing with multi-agent interactions and uncertainties.
- Introducing market share and dynamic game ideas can target the reality that retailers with large market shares are vulnerable to attacks, obtain a more accurate security investment level through multiple rounds of iterations, and enhance the model's adaptability to the competitive landscape.
- While considering security externalities, quantify the budget investment benefits and combine the Nash equilibrium mechanism to finely characterize the coordination and allocation of security resources among multiple retailers. This provides a feasible path to avoid an imbalance in security investment and efficiency loss.

This paper proposes a "nonlinear budget-constrained cybersecurity investment optimization model based on variational inequality and projection shrinkage algorithm" method. This method has the following three contributions to the game problem of supply chain cybersecurity investment under multi-retailer competition:

- It overcomes the difficulty of solving the multi-agent, non-cooperative game under budget constraints. Through the iterative convergence mechanism of the projection contraction algorithm, the solution stability is achieved under the coexistence of multi-party decision variables and constraints.
- It improves the robustness of attacking losses and market share fluctuations. In the sensitivity analysis, it is verified that this method can dynamically evaluate potential risks and investment returns, enabling enterprises to better respond to changes in demand and escalating security threats.
- It provides a highly operational quantitative basis for the security deployment of digital supply chain networks: different security means and technical investments can be flexibly embedded in the model framework to help decision-makers maximize security benefits within a limited budget.

The rest of this paper is organized in the following order: Chapter 2 explains the concepts and theories related to supply chain network security and sorts out the theoretical basis of variational inequality, projection shrinkage algorithm, and Nash equilibrium; Chapter 3 builds a supply chain network security model with nonlinear constraints on this basis and gives the corresponding solution process; Chapter 4 verifies the feasibility and effectiveness of the model through numerical simulation and sensitivity analysis; Chapter 5 summarizes the research conclusions and puts forward several suggestions for supply chain network security management.

## 2　Related Work

### 2.1　Modeling and Strategy Analysis of Supply Chain Network Security

In recent years, more and more studies have been devoted to exploring the interactive mechanism between network security investment and risk management in the supply chain environment. Various theoretical and model tools have been developed. Early representative studies focused on network equilibrium theory and variational inequality methods. For example, Nagurney et al.'s work at the Hypernetwork Research Center [2–4] and Qian et al. [5] explored the interaction between enterprise information sharing and security investment based on the Nash equilibrium model; Liu et al. [6], Ezhei et al. [7] and Solak et al. [8] further used differential games and stochastic programming techniques to study the impact of security interdependence and budget constraints on information sharing and investment decisions. This type of method can better characterize the decision-making process of multiple subjects in security protection, especially when considering partial or complete centralized decisions. It can identify the impact of externalities and coordination on security expenditures.

At the same time, some scholars focus on more specific issues such as security costs, risk compensation, and interruption effects. Li and Xu[9] emphasized the relationship between joint decision-making and third-party risk propagation in a dual-tier supply chain architecture, pointing out that an effective compensation mechanism can motivate security investment and reduce overall costs; Cheung et al.[10], Aldrighetti et al.[11], and Sawik[12] discussed the dynamic adjustment method of supply chain security strategy from the perspectives of IoT attack threats, cost analysis, and linear programming. Colajanni et al. [13] introduced the weighted average of security levels into a nonlinear budget constraint model, providing a quantitative approach to security investment under fixed demand conditions. The research methods based on game theory, network equilibrium, and multi-level optimization provide a relatively systematic theoretical framework for supply chain security investment strategies. However, under highly constrained budgets or multi-retailer competition, the problem of unbalanced security investment caused by market share differences has not received sufficient attention.

### 2.2　Integration of Supply Chain Cybersecurity and Emerging Technologies

In addition to an in-depth exploration of security investment decisions at the model level, some studies also focus on the empowerment and challenges of new technologies in supply chain cybersecurity. Wang et al. [14] focused on the network risks and security risks faced by different types of enterprises when deploying cloud computing services; Iranmanesh et al. [15] emphasized the application potential of blockchain technology in supply chain management and the importance of establishing a global unified security protocol; Akhtar [16] and Su [17] explored how enterprises should adapt to more complex cyber-attacks in the context of digital transformation from the perspectives of artificial intelligence and cross-border data flow. Similarly, in the domestic research system, Wang Mao[18], Gong Xuejiao [19], Zhang Xiaofei et al. [20] proposed security management strategies suitable for local enterprises based on variational inequalities and two-layer game models, combined with analyzing industrial Internet and information security investment restrictions; Zhang Zijian et al. [21], Xu Lu [22] and Yao Hong [23] further revealed the relationship between security externalities and budget constraints from network vulnerability, diversified risks, and adaptive strategies. Recently, Sun Jie et al. [26], Wang Lei [27], and Pan Yuhong [28] have respectively explored how digital technology, machine learning, and supply chain financial services can be combined with security management mechanisms to achieve a dual improvement in efficiency and protection capabilities.

However, Wang et al. [14], Iranmanesh et al. [15], Akhtar [16], and Su [17] mainly proposed security protection solutions from the technical aspects of cloud computing, blockchain, artificial intelligence, and cross-border data flow. Although they have achieved certain results in specific business scenarios, they mainly focus on risk control in a single dimension and fail to fully consider the imbalance of security investment caused by uneven market share among multiple retailers. Wang Mao [18], Gong Xuejiao [19], and Zhang Xiaofei et al. [20] explored information security investment restrictions and industrial Internet collaboration strategies based on variational inequalities and two-layer game models. Still, they paid more attention to the resource allocation of local enterprises in the primary digital environment. They did not further quantify the capital efficiency and security benefits when large-scale retailers were the focus of the attack. Zhang Zijian et al. [21], Xu Lu [22], and Yao Hong [23] conducted research on network vulnerability, diversified risks, and adaptive strategies and proposed to alleviate the "free riding" phenomenon between nodes through incentive or compensation mechanisms. However, there is still a lack of in-depth discussion on how large retailers with minimal budgets can dynamically balance security needs in a competitive environment. Sun Jie et al. [26], Wang Lei [27], and Pan Yuhong [28] explored how to strengthen the overall protection of the supply chain and improve decision-making efficiency from the perspective of integrating digital technology, machine learning, and supply chain financial services. However, they focus on improving the synergy between the technical and economic ends. They only briefly mentioned that companies with large market shares are prone to becoming the primary targets of attack and the security externalities caused, but they lack systematic quantitative analysis.

In summary, many researchers have carried out multi-level theoretical and applied research on supply chain cybersecurity investment and risk management, covering model methods such as multi-agent games, variational inequalities, difference games, and stochastic programming, as well as scientific and technological processes such as cloud computing, blockchain, and artificial intelligence. However, when dealing with multi-retailer scenarios with highly constrained budgets and noticeable market share differences, existing research often lacks detailed quantitative methods to characterize security investment imbalances and potential externalities. How to balance limited resources with high-security requirements has not yet been systematically solved for large retailers vulnerable to attacks. Based on this, this study proposes a nonlinear budget-constrained cybersecurity investment optimization model based on variational inequalities and projection shrinkage algorithms.

## 3   Method

### 3.1   Variable Definition

The variables and parameter symbols used in this paper are shown in **Table 1**.

**Table 1.** Symbols Description

| Symbol | Symbol Description |
|---|---|
| $m$ | Assume there are $m$ retailers, $x=1,2,...m$ |
| $n$ | Assume there are $n$ demand markets, $y=1,2,...n$ |
| $d_y$ | Demand market $y$ for the entire product, $y=1,2,...n$ |
| $B_x$ | Budget constraint of retailer $x$ in cybersecurity investment, $x=1,2,...m$ |
| $Q_{xy}$ | Number of products traded between retailer $x$ and demand market $y$, $x=1,2,...m$, $y=1,2,...n$ |
| $Q_y$ | Demand for the required products in demand market $x$, $x=1,2,...m$ |
| $u_x$ | Cybersecurity level of retailer $x$, $x=1,2,...m$ |
| $\bar{u}$ | Cybersecurity level of the supply chain network as a whole |
| $p_x$ | Probability of successful cyberattack on retailer $x$, $x=1,2,...m$ |
| $\rho_y$ | Demand price of the product in demand market $y$, $y=1,2,...n$ |
| $c_x$ | All costs incurred by retailer $x$ for the product before the transaction, $x=1,2,...m$ |
| $c_{xy}$ | Costs incurred by retailer X when trading with consumer Y, $x=1,2,...m$, $y=1,2,...n$ |
| $D_x$ | Attack losses incurred by retailer $x$ after being attacked by cyberattack, $x=1,2,...m$ |
| $t_x$ | Market share of retailer $x$, $x=1,2,...m$ |

### 3.2   Method Description

This study proposes a supply chain network security investment model with $m$ retailers as participants. Each retailer is represented by $x$, and $n$ demand markets are considered, each of which is marked by $y$. The topological structure of the supply chain network model is shown in **Fig. 1**.

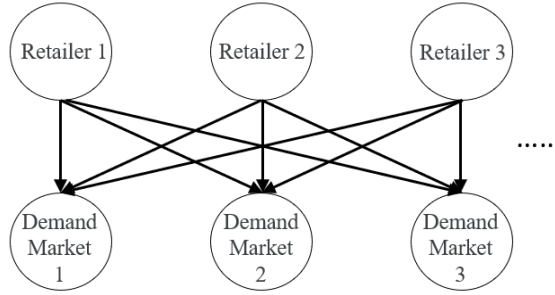

**Fig. 1.** Topological structure of the model.

For demand market $y$, total demand satisfies the following conditions:

$$d_y = \sum_{x=1}^{m} Q_{xy}, y = 1...n \tag{1}$$

$$h_x(u_x) = -\ln(1 - u_x) \tag{2}$$

The relationship between network security level and investment cost $h_x(u_x)$ is assumed to be: for the retailer $x$, $h_x(0) = 0$ means that the retailer's security level is zero and the investment is also zero; conversely, $h_x(1) = \infty$ means that the retailer's network security level is one and the investment cost reaches infinity.

To examine the impact of budget constraints on network security, we add constraints:

$$h_x(u_x) \leq B_x \tag{3}$$

The probability of retailer $x$ being attacked successfully can be expressed as:

$$p_x = (1 - u_x)(1 - \bar{u}) \tag{4}$$

Where $0 < u_x < 1, x = 1, 2...m$.

This paper refers to the models of Nagurney (2015) [3] and Nagurney (2017) [2], combining the cost of cybersecurity investment, the risk of potential cyberattacks, and the retailer's profit when not attacked to form a comprehensive economic utility framework to quantify the impact of cybersecurity investment on the retailer's expected utility. $f_x(Q,u)$ is the total profit of the retailer $x$ when no cybersecurity investment is made and no cyberattack occurs:

$$f_x(Q,u) = \sum_{y=1}^{n} \rho_y(Q,u)Q_{xy} - c_x \sum_{y=1}^{n} Q_{xy} - \sum_{y=1}^{n} c_{xy}(Q_{xy}) \tag{5}$$

Then the total expected utility $E(U_x)$ of retailer $x$ is:

$$E(U_x) = (1-p_x)f_x(Q,u) + p_x(f_x(Q,u) - D_x) - h_x(u_x)$$
$$= f_x(Q,u) - D_x p_x - h_x(u_x) \quad (6)$$

For convenience, the feasible solution set $x$ of retailers $K^x \equiv (Q_x, u_x)$ is defined, where each retailer adopts a non-cooperative competition strategy in cybersecurity investment and strives to optimize its expected profit maximization. Each retailer determines their strategy vector $(Q_x, u_x)$ to maximize the expected utility $E(U_x)$. The maximization problem is converted into a minimization problem for ease of solution. The specific form of the transformed expected utility function on the feasible solution set $K^x$ is:

$$Min\ F(X_x, \hat{X}_x^*) = -Max\ E(U_x) \quad (7)$$

Where $X_x \equiv (Q_x, u_x)$ represents the feasible solution for retailer $x$, $\hat{X}_x \equiv (X_x, ..., X_{x-1}, X_{x+1}, ..., X_m)$ represents the feasible solution for retailers other than retailer $x$, and $\hat{X}_x^*$ represents the optimal solution for retailers other than retailer $x$. The minimum value problem is simplified to the following formula:

$$Min_{(Q_x, u_x)} \hat{F}(X_x, \hat{X}_x^*) \quad (8)$$

$$G(X_x) = -\ln(1-u_x) - B_x \leq 0 \quad (9)$$

Where $G(X_x)$ represents the budget constraint of supply chain cybersecurity, the Lagrangian function of this problem is $L(X_x, \hat{X}_x^*, \lambda_x)$, and its specific expression is:

$$L(X_x, \hat{X}_x^*, \lambda_x) = \hat{F}(X_x, \hat{X}_x^*) + \lambda_x G(X_x) \quad (10)$$

$\hat{F}(X_x, \hat{X}_x^*)$ represents the expected utility function after the objective function is transformed, $\lambda_x$ is the Lagrange multiplier; $L(X_x, \hat{X}_x^*, \lambda_x)$ represents the Lagrangian function containing the constraints; the optimal solution of function $L(X_x, \hat{X}_x^*, \lambda_x)$ satisfies the following variational inequality conditions:

$$\nabla_{X_x} L(X_x^*, \hat{X}_x^*, \lambda_x^*) \times (X_x - X_x^*) + (-G(X_x^*)) \times (\lambda_x - \lambda_x^*) \geq 0, \forall (X_x, \lambda_x) \in K^x \times R_+ \quad (11)$$

Where $\nabla_{X_x} L(X_x^*, \hat{X}_x^*, \lambda_x^*)$ represents the gradient at the optimal solution, $G(X_x^*)$ represents the budget constraint of retailer $x$ at the optimal solution, $\lambda_x^*$ represents the optimal Lagrange multiplier of operator $x$, and $R_+$ represents a set of positive real numbers.

The variational inequality in this paper adopts the projection and contraction algorithm to solve the equilibrium solution [29]. Let $X \equiv (Q, u, \lambda)$ represent the feasible set of functions, $F(X) \equiv (\hat{F}^1(X), \hat{F}^2(X), \hat{F}^3(X))$, $F(X)$ is a composite vector composed of $\hat{F}^1_{xy}(X)$, $\hat{F}^2_x(X)$, and $\hat{F}^3_x(X)$, $\hat{F}^1(X)$ is the derivative of the Lagrangian function concerning the transaction volume, $\hat{F}^2(X)$ is the derivative of the Lagrangian function concerning the network security level, and $\hat{F}^3(X)$ is the derivative of the Lagrangian function concerning the Lagrangian multiplier.

$$\hat{F}^1_{xy}(X) \equiv \left[ c_x + \frac{\partial c_{xy}(Q_{xy})}{\partial Q_{xy}} - \hat{\rho}_y(Q,u) - \sum_{k=1}^{n} \frac{\partial \hat{\rho}_k(Q_x, u_x)}{\partial Q_{xy}} \times Q_{xk} \right] \quad (12)$$

$$\hat{F}^2_x(X) \equiv \left[ \frac{\partial h_x(u_x)}{\partial u_x} - (1 - \frac{1}{m}\sum_{y=1}^{m} u_x + \frac{1-u_x}{m})D_x - \sum_{k=1}^{n} \frac{\partial \hat{\rho}_k(Q,u)}{\partial u_x} \times Q_x + \frac{\lambda_x}{1-u_x} \right] \quad (13)$$

$$\hat{F}^3_x(X) \equiv -\ln(1-u_x) - B_x, \forall x \quad (14)$$

The specific steps of projection iteration are as follows:

a. Parameter initial value setting: step size $\beta_0 = 1$, the upper limit of shrinkage factor $v \in (0,1)$, the lower limit of shrinkage factor $\mu \in (0,1)$, relaxation factor $\rho = 1.9$, number of iterations $i = 0$; the initial value of feasible solution $X^0_x = (Q^0_{xy}, u^0_x, \lambda^0_x)$;

b. Projection generates estimated points:

$$\tilde{X}^i = P_K \left[ X^i - \beta_i F(X^i) \right] \quad (15)$$

The step size of the projection shrinkage $\beta_i = \frac{2}{3}\beta_i * \min\left\{1, \frac{1}{r_i}\right\}$, the shrinkage scale factor $r_i = \frac{\beta_i \| F(X^i) - F(\tilde{X}^i) \|}{\| X^i - \tilde{X}^i \|}$, $\| F(X^i) - F(\tilde{X}^i) \|$, and $\| X^i - \tilde{X}^i \|$ represent the norm of the difference between the vector $F(X^i)$ and $F(\tilde{X}^i)$ and the difference between the vector $X^i$ and the predicted point $\tilde{X}^i$, respectively; $F(X^i)$ is $F(X)$ after the $i$ th iteration, and for the estimated point $\tilde{X}^i$, the specific projection is:

$$\tilde{Q}^i_{xy} = \max\left\{0, \min\left\{\bar{Q}_{xy}, Q^i_{xy} - \beta_i \left( \left[ c_x + \frac{\partial c_{xy}(Q_{xy})}{\partial Q_{xy}} - \hat{\rho}_y(Q,u) \right] - \sum_{k=1}^{n} \frac{\partial \hat{\rho}_k(Q_x, u_x)}{\partial Q_{xy}} \times Q_{xk} \right) \right\}\right\} \quad (16)$$

$$\tilde{u}_x^i = \max\left\{0, \min\left\{\hat{u}_x, u_x^i - \beta_i \left(\begin{array}{c} \dfrac{\partial h_x(u_x)}{\partial u_x} - (1 - \dfrac{1}{m}\sum_{y=1}^{m} u_x + \dfrac{1-u_x}{m})D_x \\ -\sum_{k=1}^{n} \dfrac{\partial \hat{\rho}_k(Q,u)}{\partial u_x} \times Q_x + \dfrac{\lambda_x}{1-u_x} \end{array}\right)\right\}\right\} \quad (17)$$

$$\tilde{\lambda}_x^i = \max\left\{0, \lambda_x^i - \beta_i\left(-\ln(1-u_x) - B_x\right)\right\} \quad (18)$$

c. If $r_i > v$, repeat step (2) and iteratively calculate the estimated point $\beta_i$ and the shrinkage factor $X^i$ by adjusting the step size $r_i$ value;

d. If $r_i \leq v$, then the following calculation is performed:

$$d(X^i, \tilde{X}^i) = (X^i - \tilde{X}^i) - \beta_i(F(X^i) - (\tilde{X}^i)) \quad (19)$$

$$\delta_i = \frac{(X^i - \tilde{X}^i)^T d(X^i, \tilde{X}^i)}{\|d(X^i, \tilde{X}^i)\|^2} \quad (20)$$

$$X^{i+1} = X^i - \rho \delta_i d(X^i, \tilde{X}^i) \quad (21)$$

Where $d(X^i, \tilde{X}^i)$ is the favorable direction vector, $\delta_i$ is the step size of the contraction algorithm, and $(X^i - \tilde{X}^i)^T$ represents the transposed vector of vector $(X^i - \tilde{X}^i)$. A new iteration point $X^{i+1}$ is generated according to the correction formula.

e. If $r_i \leq \mu$, then update the value of $\beta_i$ through $\beta_i = \beta_i * 1.5$; otherwise, update the value of $\beta_i$ through $\beta_{i+1} = \beta_i$ and the value of $\sigma$ through $\sigma = \sigma + 1$, and repeat step (2) until all strategies converge, which is the optimal solution.

## 4  Numerical Simulation Results and Sensitivity Analysis

### 4.1  Simulation Experiment 1

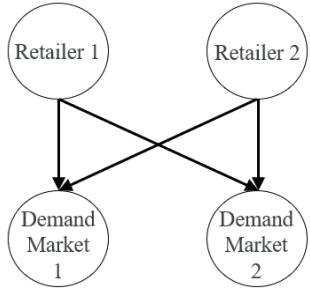

**Fig. 2.** Topology diagram of numerical simulation analysis model (Experiment 1)

This section verifies the effectiveness of the above model through numerical simulation. In Experiment 1, a supply chain network consisting of two retailers and two demand markets is constructed, and its network topology is shown in **Fig. 2**.

In this experiment, $Q_{xy}$ is initialized to 1, $0 \leq Q_{xy} \leq 100$. The security level $u_x$ of each retailer is set to 0. The budget constraint of each retailer is represented by the Lagrange multiplier $\lambda$. Assume that the probability of retailer $x$ being attacked by a cyber-attack is $p_x$, the budget constraint of retailer $x$ in cyber security investment is $B_x$, the fixed cost of retailer $x$ for the product before the transaction is $c_x$, and the variable cost of retailer $x$ when trading with consumer $y$ is $c_{xy}$. This paper corrects the attack loss $D_x$ generated by retailer M after a cyber-attack and introduces the market share $t_x$ of retailer $x$ as an elastic influencing variable. In Experiment 1, it is assumed that the market share of Retailer 1 is $t_1 = 0.76$ and the market share of Retailer 2 is $t_2 = 0.24$. The specific cost calculation formula is:

$$c_{11}(Q_{11}) = (Q_{11}^2 + 2Q_{11}) * (1+t_1) \tag{22}$$

$$c_{12}(Q_{12}) = (0.5Q_{12}^2 + 2Q_{12}) * (1+t_1) \tag{23}$$

$$c_{21}(Q_{21}) = (Q_{21}^2 + 2Q_{21}) * (1+t_2) \tag{24}$$

$$c_{22}(Q_{22}) = (0.5Q_{22}^2 + 2Q_{22}) * (1+t_2) \tag{25}$$

The price demand function for different commodities is:

$$\rho_1(d,u) = -2d_1 + 0.2\left(\frac{u_1+u_2}{2}\right) + 120 \tag{26}$$

$$\rho_2(d,u) = -d_2 + 0.4\left(\frac{u_1+u_2}{2}\right) + 250 \tag{27}$$

The cybersecurity investment budgets of Retailer 1 and Retailer 2 are:

$$B_1 = 3*(1+t_1) \tag{28}$$

$$B_2 = 3*(1+t_2) \tag{29}$$

$$h_1(u_1) = -\ln(1-u_1) \leq B_1 \tag{30}$$

$$h_2(u_2) = -\ln(1-u_2) \leq B_2 \tag{31}$$

$$p_1 = (1-u_1)(1-\bar{u})*(1+t_1) \tag{32}$$

$$p_2 = (1-u_2)(1-\bar{u})*(1+t_2) \tag{33}$$

$$c_1 = 10*(1+t_1) \tag{34}$$

$$c_2 = 10*(1+t_2) \tag{35}$$

The losses incurred by Retailer 1 and Retailer 2 after being attacked by the cyber-attack are:

$$D_1 = 100*(1+t_1) \tag{36}$$

$$D_2 = 100*(1+t_2) \tag{37}$$

In summary, the profit function $f_x(Q,u)$ and expected utility function $E(U_x)$ of Retailer 1 and Retailer 2 can be expressed as:

$$f_1(Q,u) = \sum_{y=1}^{n} \rho_y(Q,u)Q_{1y} - c_1 \sum_{y=1}^{n} Q_{1y} - \sum_{y=1}^{n} c_{1y}(Q_{1y}) \tag{38}$$

$$f_2(Q,u) = \sum_{y=1}^{n} \rho_y(Q,u)Q_{2y} - c_2 \sum_{y=1}^{n} Q_{2y} - \sum_{y=1}^{n} c_{2y}(Q_{2y}) \tag{39}$$

$$E(U_1) = f_1(Q,u) - D_1 p_1 - h_1(u_1) \tag{40}$$

$$E(U_2) = f_2(Q,u) - D_2 p_2 - h_2(u_2) \tag{41}$$

Retailer 1's feasible solution set $K^1 \equiv (Q_1, u_1)$ is dedicated to maximizing the expected utility $E(U_1)$ in cybersecurity investment. The maximization problem is converted into a minimization problem. The specific form of the transformed expected utility function on the feasible solution set $K^1$ is:

$$Min\ F(X_1, \hat{X}_1^*) = -Max\ E(U_1) \tag{42}$$

Where $X_1 \equiv (Q_1, u_1)$ is the feasible solution for Retailer 1, $\hat{X}_1 \equiv X_2$ is the feasible solution for Retailer 2, and $\hat{X}_1^*$ is the optimal solution for Retailer 2. The minimum problem is simplified to:

$$Min_{(Q_1, u_1)} \hat{F}(X_1, \hat{X}_1^*) \tag{43}$$

$$G(X_1) = -\ln(1-u_1) - B_1 \leq 0 \tag{44}$$

$G(X_1)$ is the network security budget constraint of Retailer 1, and the same is true for Retailer 2. The Lagrangian function of Retailer 1 is $L(X_1, \hat{X}_1^*, \lambda_1)$, and the specific expression is:

$$L(X_1, \hat{X}_1^*, \lambda_1) = \hat{F}(X_1, \hat{X}_1^*) + \lambda_1 G(X_1) \tag{45}$$

$\lambda_1$ is the Lagrange multiplier of Retailer 1, and the optimal solution of function $L(X_1, \hat{X}_1^*, \lambda_1)$ satisfies the variational inequality condition. Composite vectors $F(X) \equiv (\hat{F}^1(X), \hat{F}^2(X), \hat{F}^3(X))$ and $X \equiv (Q, u, \lambda)$ are the feasible sets of functions; $\hat{F}^1(X)$ is the derivative of the Lagrange function concerning the transaction volume, $\hat{F}^2(X)$ is the derivative of the Lagrange function concerning the network security level, and $\hat{F}^3(X)$ is the derivative of the Lagrange function concerning the Lagrange multiplier.

$\tilde{X}^i$ is the estimated point of the projection shrinkage algorithm. For Retailer 1, the specific projection formula is as follows:

$$\tilde{Q}_{1y}^i = \max\left\{0, \min\left\{\hat{Q}_{1y}, Q_{1y}^i - \beta_i \left(\begin{bmatrix} c_1 + \frac{\partial c_{1y}(Q_{1y})}{\partial Q_{1y}} - \hat{\rho}_y(Q, u) \\ -\sum_{k=1}^{n} \frac{\partial \hat{\rho}_k(Q_1, u_1)}{\partial Q_{1y}} \times Q_{1k} \end{bmatrix}\right)\right\}\right\} \quad (46)$$

$$\tilde{u}_1^i = \max\left\{0, \min\left\{\hat{u}_1, u_1^i - \beta_i \left(\begin{array}{c} \frac{\partial h_1(u_1)}{\partial u_1} - (1 - \frac{1}{m}\sum_{y=1}^{m} u_x + \frac{1-u_x}{m})D_1 \\ -\sum_{k=1}^{n} \frac{\partial \hat{\rho}_k(Q, u)}{\partial u_1} \times Q_1 + \frac{\lambda_1}{1-u_1} \end{array}\right)\right\}\right\} \quad (47)$$

$$\tilde{\lambda}_1^i = \max\left\{0, \lambda_1^i - \beta_i \left(-\ln(1-u_1) - B_1\right)\right\} \quad (48)$$

The iteration point is corrected by adjusting the step size and the shrinkage factor $r_i$ until Retailer 1's and Retailer 2's expected profits converge, which is the optimal solution to this problem.

This paper considers two retailers to be optimal simultaneously under interaction, so the model uses the alternating optimization equilibrium model; that is, each participant optimizes its strategy based on the current strategy of other participants. This process is iterative, and each round of iteration is based on the best response to others' strategies in the previous round. After alternating optimization, the final utilities of the two retailers are obtained $E(U_1)$, $E(U_1)$, transaction volumes $Q_{11}$, $Q_{12}$, $Q_{21}$, $Q_{22}$ and network security levels $u_1$, $u_2$.

The algorithm was programmed and implemented in the MATLAB (R2021 a) environment, and the simulation results are shown in **Table 2**.

**Table 2.** Summary of simulation experiment results.

| $Q_{11}$ | $Q_{12}$ | $Q_{21}$ | $Q_{22}$ | $u_1$ | $u_2$ |
|---|---|---|---|---|---|
| 10.94 | 30.25 | 11.78 | 31.73 | 0.96 | 0.95 |

As shown in **Table 2**, the security investment levels of the two retailers are 0.96 and 0.95, respectively. The sales volume of Retailer 1 is 10.94 and 30.25, while the sales volume of Retailer 2 is 11.78 and 31.73. The transaction volumes of the two retailers are similar. Retail enterprises have chosen a higher level of cybersecurity investment in this environment because enterprises realize that maintaining a high level of cybersecurity is critical in an increasingly digital and networked business environment. A higher level of cybersecurity investment has many practical significances: in the cost-benefit trade-off, although maintaining a high level of cybersecurity requires significant investment, compared with the enormous losses that cyber-attacks may cause, such investment is reasonable and necessary. Retailers' high-security investment can be considered a preventive measure against potential losses, especially when facing high-probability cyber-attacks. In addition, consumers' trust in enterprises depends significantly on the security of their data. A high level of security investment prevents data leakage and enhances consumers' trust and loyalty to the brand, thereby enhancing competitive advantages in the market.

### 4.2  The Impact of Budget Constraints on Cybersecurity

Experiment 2 is conducted based on Experiment 1 to analyze the impact of budget constraints on network security. Keeping other conditions unchanged, Experiment 2 changes Retailer 1's budget $B_1$ from 2.00 to 3.50 and obtains the changing trends of $B_1$ and $u_1$ under different $u_1$, as shown in **Fig. 3**.

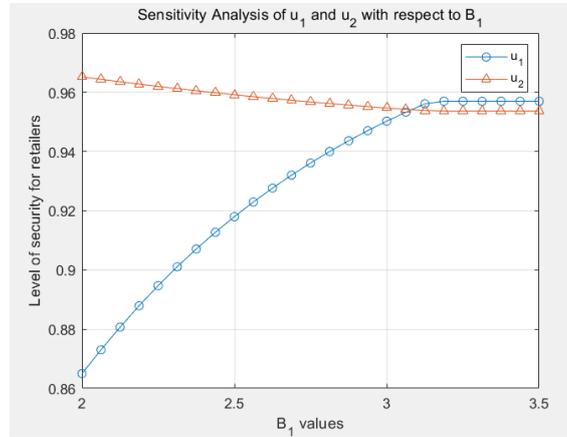

**Fig. 3.** The impact of B1 on network security level.

As shown in **Fig. 3**, the network security level of Retailer 1 $u_1$ shows a gradual upward trend with the increase in budget, showing a strong positive correlation. When $B_1$ after 3.06, the network security level of Retailer 1 $u_1$ is higher than that of Retailer 2 $u_2$. When $B_1$ continues to increase to 3.50, the growth trend slows down and gradually becomes stable. This shows that with the increase in investment, retailers have reached the diminishing marginal utility of investment in network security, and further increasing the budget will not bring the same proportion of security level improvement. This conclusion is similar to the results of the literature in the paper [30]. The network security level of Retailer 2 $u_2$ is relatively stable within the entire range of $B_1$ and shows a slight downward trend.

Through the dynamic analysis of the cybersecurity levels of Retailer 1 and Retailer 2 under different budget constraints, security investment requires sophisticated budget control and wise strategic planning. Determining the optimal security investment level can analyze the cost-effectiveness. By comparing the losses caused by potential security incidents with the security investment costs, it is necessary to determine the security measures to be taken and the reasonable scale of investment. At the same time, enterprises can also sort cybersecurity investment projects based on risk identification and risk assessment results and prioritize investing in security threats that have the most significant impact on the business to ensure that cybersecurity investment can meet security needs and economic benefits. This may involve adjusting the budget to ensure its cybersecurity investment is economically beneficial and has the necessary flexibility to cope with the rapidly changing technological environment and market demand. In practice, enterprises can set up flexible budget management mechanisms to adjust to changes in security threats and the actual results of projects during the fiscal year to improve business efficiency and competitiveness.

### 4.3 Impact of Attack Losses on Network Security

Based on Experiment 1, Experiment 3 was conducted to analyze the impact of attack losses on network security. While keeping other conditions unchanged, Experiment 3 changed Retailer 1's $D_1$ loss from 120.00 to 200.00 and obtained the changing trends of $D_1$ and $u_1$ under different $u_2$, as shown in **Fig. 4**.

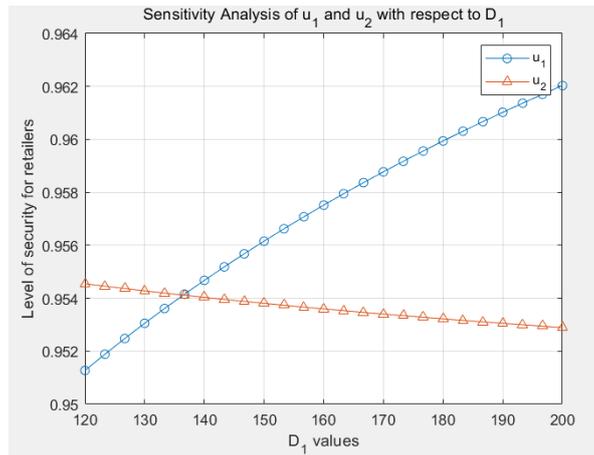

**Fig. 4.** Impact of D1 on network security level.

As shown in **Fig. 4**, the cybersecurity level of Retailer 1 shows a significant upward trend. When $D_1 = 120$, $u_1$ is approximately 0.951, and when $D_2 = 200$, $u_1$ is close to 0.962. This trend reflects that Retailer 1 pays more attention to cybersecurity as the potential loss increases, and its cybersecurity level will gradually improve. Retailer 2's cybersecurity investment decision is somewhat affected by Retailer 1, and Retailer 2's cybersecurity level shows a slow downward trend. When the loss value of retailer 1 $D_1$ is 137, the cybersecurity levels of the two retailers are equal. This means that under this loss risk, the two retailers adjust their cybersecurity level investment strategies to the same level.

The losses caused by cyber-attacks on enterprises can be direct financial or indirect losses such as brand reputation and customer trust. The cybersecurity level of Retailer 1 $u_1$ increases significantly with the increase of $D_1$, which shows that Retailer 1 is willing to increase cybersecurity investment when facing increased potential losses, which reflects its intense sensitivity to risk response and loss avoidance. This is consistent with risk management theory; enterprises tend to strengthen preventive measures when facing more significant risks. As potential losses increase, enterprises' risk awareness increases. This, in turn, increases cybersecurity budgets; enterprises are willing to invest in more advanced security technologies and services to defend against possible cyber-attacks. This response to high risks and loss potential is particularly evident in enterprises that handle sensitive information or are in high-risk industries (such as financial services and healthcare). Therefore, enterprises should strengthen the implementation and supervision of cybersecurity policies and processes, establish and enforce strict access control policies, and ensure that only authorized personnel can access sensitive information and systems. At the same time, enterprises should establish a comprehensive security monitoring system to detect and respond to potential security threats promptly by continuously monitoring network traffic, system logs, and abnormal activities, thereby reducing the losses caused by cyber-attacks and adapting to the ever-changing threat environment.

### 4.4 The Impact of Market Share on Cybersecurity

Based on Experiment 1, Experiment 4 was conducted to analyze the impact of market share on network security. While keeping other conditions unchanged, Experiment 4 changed the market share of Retailer 1 $t_1$ from 0.55 to 0.89 and obtained the changing trends of $t_1$ and $u_1$ under different $u_2$, as shown in **Fig. 5**.

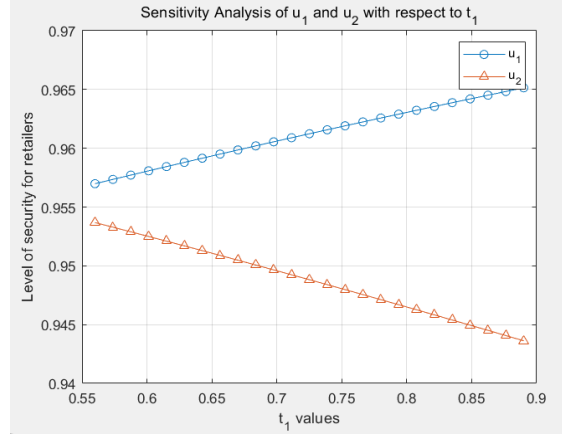

**Fig. 5.** The impact of t1 on network security level.

**Fig. 5** shows the trend of the cybersecurity levels $u_1$ and $u_2$ of Retailer 1 and Retailer 2 as the market share $t_1$ of Retailer 1 changes. The cybersecurity level $u_1$ of Retailer 1 increases significantly as the market share increases. This indicates that as the market share of Retailer 1 increases, it is more vulnerable to cyber-attacks, so it tends to increase investment in cybersecurity. The cybersecurity level $u_2$ of Retailer 2 decreases gradually as the market share $t_1$ of Retailer 1 increases. This may be because as the market share of the competitor Retailer 1 increases, the market share of Retailer 2 decreases relatively, thereby reducing investment in cybersecurity and resulting in a lower cybersecurity level. When a company faces a decline in its market position in the real business environment, it may reduce investment in non-directly profitable areas, including cybersecurity investment.

In general, market share has a positive impact on cybersecurity investment. The increase in retailer market share will improve its cybersecurity level and affect the cybersecurity strategies of competitors in the entire market. Retailers with a larger market share, to maintain customer trust and corporate reputation, are more inclined to increase cybersecurity investment and improve cybersecurity levels. When the market expands, or the market share increases, companies should foresee the corresponding increase in cyber-attack risks and take measures to enhance cybersecurity protection. For example, measures such as strengthening intrusion detection systems, implementing advanced endpoint protection, and strengthening employee security training can be taken to deal with increased risks. Retailers with a smaller market share control operating costs and reduce cybersecurity investment. Companies must scientifically evaluate their market conditions and adjust cybersecurity investment according to their business goals and external threats.

## 4.5 The Impact of The Number of Retailers on Cybersecurity

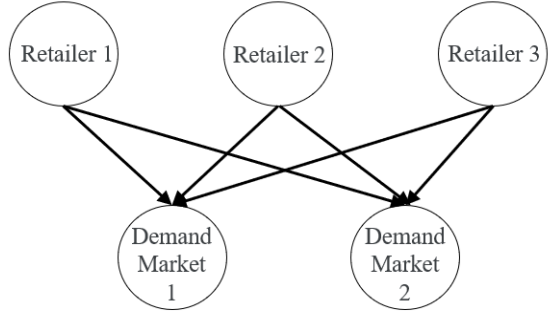

**Fig. 6** Topological structure of Experiment 5

Based on Experiment 1, Experiment 5 adds Retailer 3 as a new entrant in the industry, assuming that the market share of each retailer is $t_1 = 0.71, t_2 = 0.20, t_3 = 0.09$. The cost and budget constraint formula for the newly added Retailer 3 is as follows:

$$c_{31}(Q_{31}) = (Q_{31}^2 + 2Q_{31}) * (1 + t_3) \tag{49}$$

$$c_{32}(Q_{32}) = (0.5Q_{32}^2 + 2Q_{32}) * (1 + t_3) \tag{50}$$

$$B_3 = 3 * (1 + t_3) \tag{51}$$

$$h_3(u_3) = -\ln(1 - u_3) \leq B_3 \tag{52}$$

$$p_3 = (1 - u_3)(1 - \bar{u}) * (1 + t_3) \tag{53}$$

$$c_3 = 10 * (1 + t_3) \tag{54}$$

$$D_3 = 100 * (1 + t_3) \tag{55}$$

The other calculation formulas are consistent with those in Experiment 1. The network security levels are analyzed as follows: $u_1 = 0.55, u_2 = 0.58, u_3 = 0.59$, and the comparative experimental results are shown in **Table 3**.

**Table 3.** Comparison of the results of Experiment 1 and Experiment 5

|       | Experiment 1 | Experiment 5 |
|-------|--------------|--------------|
| $u_1$ | 0.96         | 0.55         |
| $u_2$ | 0.95         | 0.58         |
| $u_3$ | \ | 0.59 |
| $\bar{u}$ | 0.955 | 0.573 |

In Experiment 5, a third retailer was added as a new entrant, which intensified market competition, and retailers paid more attention to cost control. The experimental results show that the investment of the original Retailers 1 and 2 in network security has decreased significantly, with the network security level of Retailer 1 decreasing from 0.96 to 0.55 and the network security level of Retailer 2 decreasing from 0.95 to 0.58. This phenomenon shows that the entry of new competitors will change existing companies' network security investment strategy. Retailers 1 and 2 may reduce their investment in network security because they need to compete with new entrants in terms of price or other marketing strategies.

The lemon effect theory in economics explains the impact of the new entrant (Retailer 3) on the market and the network security level of existing competitors (Retailers 1 and 2). The lemon effect was proposed by economist George Akerlof in 1970. It refers to the impact of the existence and spread of low-quality products or services on the entire market in a market economy. It mainly describes the problem of market mechanism failure under information asymmetry. The entry of new entrants has intensified market competition, causing existing retailers to reduce cybersecurity investment to control costs. Due to information asymmetry, this behavior will affect the decision-making of other competitors, forming a lemon effect within the same industry. At the same time, when some new entrants take the initiative to reduce cybersecurity investment to save costs, existing retailers may also be forced to follow suit to maintain competitiveness in terms of price, etc., which may, in turn, reduce the average level of cybersecurity in the entire industry, thus forming a lemon effect phenomenon in the level of cybersecurity. Therefore, to reduce the lemon effect within the same industry, retailers need to increase the transparency of cybersecurity investment so that competitors can better understand each other's actual situation. One feasible way is to establish a trust mechanism within the industry through third-party cybersecurity certification and information disclosure. In addition, the government and industry organizations can also formulate unified cybersecurity standards and disclosure systems to regulate the cybersecurity practices of enterprises in the industry, thereby reducing information asymmetry and curbing the lemon effect within the same industry. Therefore, retailers need to increase the transparency of cybersecurity investment, and the government and industry organizations should also formulate relevant standards and systems to alleviate the negative impact of the lemon effect within the same industry and promote the improvement of the cybersecurity level of the entire industry.

### 4.6 Supply Chain Cybersecurity Investment Advice

Based on the above model simulation results and sensitivity, this study proposes the following supply chain cybersecurity investment strategy recommendations:

a) **Dynamic risk assessment and flexible strategy adjustment.** Enterprises must maintain a dynamic market risk assessment to maintain enterprise cybersecurity. Budget constraints force retailers to weigh the costs and benefits of cybersecurity investment, including initial investment and subsequent maintenance costs.

Therefore, financial planning must be flexibly adjusted to cope with the uncertainty of attack losses and changing security threats. For example, with the evolution of cloud computing and Internet of Things technologies, retailers must quickly adjust security strategies to protect data and corporate assets. With increasing attack losses, enterprises should take proactive and preventive measures, such as implementing multi-factor authentication and zero-trust network architecture to guard against unknown threats. These measures reduce direct financial losses and help avoid reputation damage caused by data leakage or system outages. Through sensitivity analysis, retailers can better understand the impact of budget and attack losses, formulate security strategies that are both economical and adaptable, protect their own and customer data, and maintain competitiveness in complex network environments.

b) **Strengthen network protection for large enterprises.** For enterprises with significant market share, their extensive business and large user base make them the main cybercrime targets. Therefore, these companies must demonstrate high vigilance and enhanced defense capabilities that match their market position. This includes adopting the most advanced security technologies, such as enhanced encryption measures, multi-factor authentication systems, and end-to-end data protection technologies, as well as implementing strict access control and data monitoring strategies. As technology and the market change, companies must continuously monitor network boundaries and use automated threat detection tools and behavioral analysis systems to identify and respond to security threats quickly. At the same time, companies should pay attention to the security dynamics of competitors, including security investments and policy updates, to ensure that they do not lag in technology and strategy. For example, companies can enhance their security frameworks by introducing advanced security technologies that competitors adopt. In addition, companies should use technological advances to optimize security operations, including automated security monitoring and response processes, to improve defense efficiency and accuracy, ensure the efficiency and economy of security measures, and adapt to the evolving market and technological environment.

c) **Promote third-party security certification.** To solve the lemon effect revealed in the experiment, especially the information asymmetry problem in the cybersecurity market, it is imperative to establish and promote a third-party certification system. Independent assessment bodies perform these certifications by international or industry standards. Examples include ISO/IEC 27001, which evaluates a company's information security management system (ISMS). Once a company obtains these certifications, it can prove to consumers the effectiveness and reliability of its network security measures. This helps consumers identify companies that implement network security investments according to high standards and point out those that do not meet the standards, effectively reducing information asymmetry in the market. The government can also promote the widespread adoption and application of such certification systems by providing policy incentives, such as tax breaks and pre-emptive purchase rights, to encourage companies to obtain and maintain these certifications. This move will enable companies to invest in security technologies and measures that meet or exceed these certification standards, thereby improving the industry's overall level of network security.

## 5 Research Conclusions and Prospects

### 5.1 Research Conclusions

In the current digital era, enterprises' cybersecurity challenges are becoming increasingly severe. Data leakage and theft bring direct financial losses and seriously damage enterprises' reputations and customer trust. Given the frequent occurrence of cybercrime incidents, finding and implementing the optimal cybersecurity strategy is particularly urgent. This article aims to provide scientific cybersecurity investment support for supply chain enterprises by analyzing multiple key factors affecting cybersecurity investment, including budget constraints, attack losses, market share changes, and new competitors.

Experiment 1 explored the direct impact of budget size on network security level and found that an increased budget can effectively improve security protection capabilities. Still, there is a phenomenon of diminishing marginal utility. Experiment 2 further analyzed the impact of attack losses on corporate security investment. The results showed that companies tend to increase security investment as potential attack losses increase to avoid more significant risks. Experiment 3 focused on the impact of market share changes on network security investment and found that increasing market share would prompt companies to improve their network security level to maintain customer trust and corporate reputation. Experiment 4 analyzed the impact of new competitors on existing companies' network security investment strategy and found that the emergence of new competitors would lead to the emergence of the lemon effect. On this basis, supply chain network security investment recommendations were put forward. This study provides a new perspective for the theoretical analysis of network security investment. It provides a decision-making reference for companies to formulate effective network security investment strategies in a complex and changing market environment.

### 5.2 Research Conclusions

Although this study analyzes the impact of variables such as budget constraints, attack losses, and market share on supply chain cybersecurity investment strategies, technological advances, changes in consumer behavior, etc., may also impact supply chain cybersecurity decisions, which need to be explored in future studies.

In addition, experimental methodology and research depth, although experimental design reveals essential trends, may not be sufficient in simulating real supply chain environments and consumer behaviors. Future research can verify and deepen the understanding of cybersecurity investment strategies by introducing machine learning and oversized data analysis methods to complex simulated transaction environments, field experiments, or empirical analysis of actual operating data to reveal complex market dynamics and consumer behavior patterns and provide a more precise theoretical basis and practical guidance for optimizing supply chain cybersecurity investment.